# Simulation of Pedestrian Movements on Slope Using Fine Grid Cellular Automata

**Dr. Siamak Sarmady**
Department of IT and Computer Engineering
Urmia University of Tech., Urmia, Iran
Email: siamaksarmady@uut.ac.ir

**Associate Professor Dr. Fazilah Haron**
College of the Computer Science and Engineering
Taibah University, Madinah, Saudi Arabia
Email: fabdulhamid@taibahu.edu.sa

**Professor Dr. Abdullah Zawawi Talib**
School of Computer Science
Universiti Sains Malaysia, Penang, Malaysia, 11800
Email: azht@usm.my

**ABSTRACT**

Cellular automata crowd simulation models have been extensively used to determine the performance and the safety of crowded structures. In conventional cellular automata systems, all the cells are of the size of a typical pedestrian. The free flow speed of a pedestrian is around 1.3m/s, and the typical size of a cell in conventional CA models is around 45cm. In order to maintain the upper boundary of movement speed, pedestrians should move 0, 1, 2 or 3 cells at each second in the simulation. These limited number of speed values pose a limitation on the accuracy of the simulated speed of pedestrians. Movement speed of pedestrians on a slope depends on the angel of the slope. The conventional CA limitations do not allow accurate simulation of pedestrian movements on slopes with different angels. Fine grid cellular automata model uses much smaller cells (e.g. 5cm or smaller). The model therefore allows much more possible speed values. In this article, fine grid cellular automata model is used to model and simulate the movements of pedestrians on a slope. A walkway scenario is used for the evaluation of the model.

## 1   Introduction

Crowd simulation models are used to estimate the performance of crowd structures such as buildings. The results of the simulations are used to increase crowd safety, comfort and performance and therefore the accuracy of the results is important. Most of the existing models do not consider the slope of the paths in the crowd system. This could affect the accuracy of the results.

Cellular automata models are among the popular methods used for crowd simulation because of their flexibility and simple and light implementation. Simple transition rules are used to determine which neighboring cell is selected by each agent for its next move. Most of the existing cellular automata models divide the simulation area into cells of the size of a pedestrian which is around 45 cm [1-5]. The relatively large size of the cells, limit the choice of movement speed for the pedestrians. Pedestrians can move 0, 1, 2 or 3 cells in each second (i.e. equal to 0, 0.45, 0.9, 1.35 m/s). The speed of pedestrians on slope depends on the free flow speed of the specific pedestrian and the angle of the slope. The limited number of speed values affects the accuracy of the models especially those which include slope in their movement paths.

Fine grid cellular automata model uses smaller cells in which pedestrians and other moving objects occupy several cells (e.g. 5 cm by 5 cm cells). The smaller cells allow to have much more speed options. Movement speed can therefore be adjusted more accurately. However, the transition rules are still simple and the model is fast and can be used for a simulation involving very large number of pedestrians. The model also allows to adjust the density-speed relation to any of the available empirical speed-density graphs [6, 7]. In this article we propose an extension to the fine grid cellular automata method which will allow to simulate slopes of different degrees and orientations in different sections of a crowd simulation model.

The paper is organized as follows. Review of the existing related work has been done in Section 2. Section 3 introduces the fine grid slope model. Section 4 presents the evaluation of the simulation model and the results. Section 5 concludes the paper.

## 2   Related Work

Crowd simulation models could be categorized into three main categories namely microscopic, mesoscopic and macroscopic. In microscopic models, the behavior of the crowd is predicted by simulating the movements and actions of the individual pedestrians. These models allow to observe and study the interactions of pedestrians with obstacles and building structures in normal and evacuation situations. These models need considerable amount of calculations for larger crowds but a typical modern desktop PC is sufficient to run them.

Mesoscopic models on the other hand, do not model the movements of individual pedestrians in the crowd. They just calculate the flow in different parts of the crowd network comprising of nodes (rooms) and links(corridors). Discrete event simulation and queueing networks are used for the calculation of the flow. These models could also determine the congestion and queueing level at different doorways and corridors in the simulation. An example of these models has been built by Lovas [8]. The model simulates the way selection behavior of pedestrians and calculates the flow of each path using the information. Another model by Hanicsh et al. could provide additional information such as average flow, average speed and average density in each region of the model [9].

In macroscopic models, crowd is considered as a liquid-like matter with different densities and speeds in different parts of it. The advantage of these models is the low computation requirements for the calculation of flow, speed and density using differential equations [10-15].

Most of the existing crowd simulation models fall in the microscopic category. Three major approaches have been adopted in this category of models namely, cellular automata, physics based and rule-based models. Cellular automata approach divides the movement space into cells and uses simple cellular automata transition rules [1-5], distance maps [16-18] or similar techniques to simulate the movements of the pedestrians between the cells. There are three types of cell-based models. In the first type of these models, just one pedestrian occupies each cell and the size of each cell is about the size of a typical pedestrian. Collision avoidance is automatically applied in these models. The second type of cell-based models allow more than one pedestrian to occupy each cell [19, 20]. These models have similarities to mesoscopic models. They provide less information about the movements of individual pedestrians but more information about subsections of the movement area in comparison to mesoscopic models. In the third type of these models, each individual pedestrian occupies more than one cell. Fine grid cellular automata [7] and the model proposed by Kirchner et al. [21] are of this type of cell-based models. In Kirchner's model pedestrians occupy 2 x 2 cells while in fine grid cellular automata cells could be much smaller (e.g. 5cm x 5cm). These models fill the gap between traditional cellular automata models and continuous models. The transition rules in cellular automata models are simple logical or mathematical rules that are evaluated independently for individual pedestrians. As a result, these models are fast and can handle large number of pedestrians and moving objects. Models by Nagel[22], Biham [23], Nagatani [24], Dijkstra [3], Blue [1], Kirchner et al.[2, 4, 5] use cellular automata or similar cell-based methods. These models mostly differ in their next cell selection algorithms. A popular method is to assign a probability to each of the neighboring (and the existing) cells based on their distance to the target and select one of them based on the probabilities. Another selection method is using precalculated distance maps [18].

The next approach in microscopic model category is Physics-based approach. The models using this approach use physical rules to determine the movements of pedestrians and their interactions with other pedestrians, building structures and the movement target. Some of the more popular models in this approach include the social forces [25], magnetic force method [26], forces model [27] and velocity obstacle model [28-32]. In social forces model imaginary repelling forces between pedestrians and obstacles (e.g. walls), repelling forces between pedestrians and other pedestrians and attraction forces between pedestrians and their targets are used to model their movements[25].

The last approach in the microscopic model category is the rule-based approach. Rule-based models use a few simple behavioral rules to model the movements of groups of creatures like birds and fishes. Reynolds' rule-based model is one of these models. It uses three rules namely, velocity matching (i.e. with other flock members), collision avoidance and flock centering (i.e. staying near other members) [33]. Rule-based models have difficulty reproducing realistic simulations of phenomena like collision avoidance, density effects and specific behaviors of pedestrians in narrow passages.

The cellular automata models reviewed in this section do not support movements of crowd on slope. In this study, cellular automata model has been improved to model the movements of pedestrians on slope and therefore provide more accurate simulations in scenarios involving sloped surfaces.

## 3 Models

The aim of this research work is to improve and extend fine grid cellular automata model [7, 34] to allow simulation of the movements of pedestrians and other moving objects on flat and sloped floors. An intentions and actions layer uses discrete-event methods to simulate the actions of moving objects (i.e. pedestrians, wheelchairs and etc.). A macroscopic movements layer simulates the way-finding behaviours and a microscopic movements layer uses the modified fine grid cellular automata model to simulate the small-scale movements among the cells.

### 3.1 Probability Model for Cellular Automata Transitions

Transition probability $P_i$ is the probability of the center point of the pedestrian map moving into neighbor $I$. It is calculated using the simplest method proposed in the fine grid model. For each neighbor $i$, the parameter $M_i$ is calculated using the following equation:

$$M_i = \frac{n_i}{R_i} \quad (3)$$

$$n_i \in \{0,1\}, \quad R_i \neq 0$$

In Equation **Error! Reference source not found.**, $R_i$ is the distance between cell $i$ and the target. The parameter $n_i$ is the collision avoidance term, and it is 0 if the neighbor i is occupied and 1 otherwise. The cell nearest to the target gets a larger $M_i$ than the others. In the next step, the neighbors are sorted according to their $M_i$ value in descending order. Given that $M_i$ is directly proportional to the desirability of a cell, the cell with the highest $M_i$ and superior rank (i.e. smaller index) should be selected most of the time. The Poisson statistical distribution (with a $\lambda \leq 1$) is used to randomly select the next cell. A softmax function can also be used to calculate the probabilities but it will possibly be more expensive considering that it is called millions of times.

As mentioned in the conditions, the equation 3 is only used when $R_i \neq 0$. $R_i = 0$ means that the neighbor $i$ is the target. In that case, cell $i$ gets a probability of 1 if it is not occupied. All other cells get a probability of 0, and the pedestrian moves into the cell which is on the target area.

In conventional cellular automata models, pedestrians will always move with their free flow speed unless they encounter an obstacle or another pedestrian. In order to produce more realistic behavior, fine grid cellular automata, calculates the density of pedestrians and other moving entities in the perception area (towards the movement direction) and uses the available empirical graphs to obtain the speed with which the pedestrian or other moving entities should move. It is then attempted to achieve that speed for the specific moving entity. Further details on the method could be obtained from the original articles [7, 34].

### 3.2 Slope and Steps

As mentioned earlier, walking speed on slope and steps depends on the slope degree and the free flow speed of the specific pedestrian. Using smaller cells allows to have more speed values and therefore easier adjustment of speed for pedestrians walking on slopes with different angle.

Pingel et al. [35] have provided a diagram depicting the relation between different slope angles and average pedestrian walking speed (Figure 1). The speed at 0º slope is around 5.11 KPH ($\simeq$1.42 m/s) which is near to the average free flow speed mentioned in other sources (e.g. Fruin's[36] and Weidmann's[37] data). The peak speed happens at around -3.1º slope (slide) but the speed drops at more steep slides. That's most likely because people tend to control their speed on steep slides to avoid falling. In positive slopes the speed drops because walking needs more energy and at steep slopes of around 40º the movement stalls.

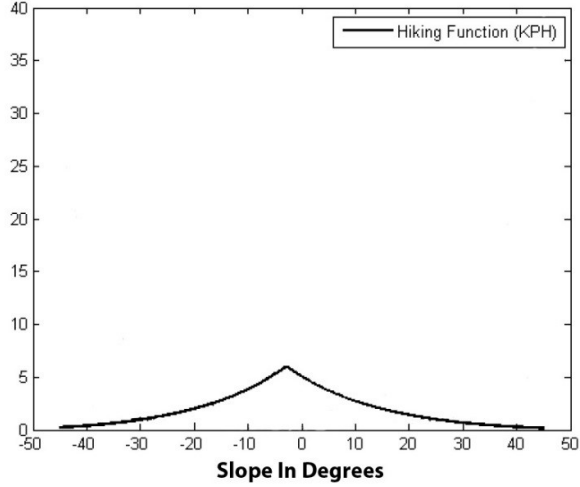

Figure 1: Hiking function showing average walking speed in relation to slope [38]

It should be noticed that the graph shows the average speed of large number of samples (e.g. pedestrians). Pedestrians may have a free flow speed that is slightly different from this graph. That is either because we use a specific fundamental density-speed graph (e.g. Fruin or Weidmann) with a different free flow speed, or because the pedestrian speeds are slightly different and determined randomly from a normal distribution (i.e. around the selected free flow speed). In our implementation, both options can be used.

In order to calculate the desired speed of each pedestrian at a specific angle, the ratio between the free flow speed of the hiking diagram (i.e. at 0º slope) to the speed at the desired slope angle is specified (Equation 1). The slope speed modifier $K_V$ is then used to scale and adjust the speed of individual pedestrians for the desired angle.

$$K_v(\Theta) = \frac{V_\Theta}{V_{0°}} \quad (1)$$

In Equation 1, $V_\Theta$ is the speed value on the hiking speed graph for a slope of $\Theta°$ and $V_{0°}$ is the speed for a 0º slope on the same graph (i.e. $\simeq 1.42$ m/s). The slope speed modifier is used in conjunction with the desired density-speed model. The speed of the pedestrian is determined from the desired fundamental density-speed diagram (i.e. Fruin or Weidmann) based on the precepted density for each pedestrian and the number is multiplied by the slope speed modifier $K_V$.

For each of the cells in the fine grid cellular automata, the slope and the direction of the slope (both in degrees) are stored in the metadata of each cell. In run time depending on the direction of pedestrian's movement; the effective slope is calculated for each individual. Based on the effective slope, the slope speed modifier $K_V$ is calculated. The effective slope ($\Theta$) is calculated from the following formula:

$$\Theta = \Theta_c \, Cos(\alpha) \quad (2)$$

where $\Theta_C$ is the slope of the cell and $\alpha$ is the angle between slope direction in the cell and the overall movement direction of the pedestrian towards the target (Figure 2).

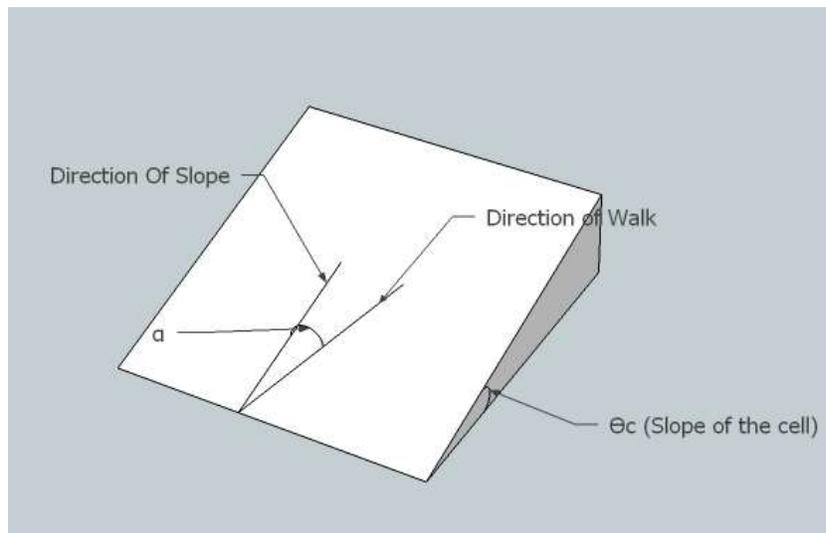

Figure 2: Direction of slope and direction of movement

If the slope of the cell and the movement direction are the same, the Cos (α) =1 and the effective slope will be equal to the cell slope. However, if the movement direction is different, the effective slope will be different. For steps, an empirical graph similar to Figure should be provided to the program. The program will use the data to calculate the speeds of pedestrian in steps with different slopes. The graph for the steps could presumably be similar to the hiking function (Figure ).

Figure shows the pseudo-code for the calculation of desired speed and the binomial probability used to allow or bar each move. The desired speed is first calculated for each pedestrian by passing the density value obtained from the pedestrian perception to a lookup function. The lookup function finds a matching speed from an empirical graph (e.g. Fruin, Weidmann, Predtechensky-Milinsky). The slope slowdown is also calculated using the Hiking function as described earlier.

```
desiredSpeed = slopeSpeedModifier * lookUpDensitySpeed(pathPedestrianDensity(), FundDiagram)

In each speed control period (eg. second):
    remainingDisplacement = desiredSpeed - performedDisplacement;

    maxAllowedMoves = remainingDisplacement / CellSize;
    binomialProbability = maxAllowedMoves / remainingStepsInPeriod;
```

Figure 3: Pseudo-code for the calculation of the binomial probability of each move

## 4  Simulation Results

The hiking function which is used as the base of the slope model is obtained from very low-density crowds so it generally depicts the free flow speed on different degrees of slope. According to the graph the speed on a flat path is around 1.42 m/s. This value is different for the free flow speed on Weidmann's empirical graph (1.344 m/s on 0-degree slope). For the experiments, two separate unidirectional walkway simulations were performed, one with a free flow speed equal to the free flow speed of hiking function and another with the free flow speed of Weidmann's graph. The results of the two simulations were then compared to the empirical values of the hiking function (Table ).

The observed densities are around 0.1 pedestrian/m$^2$ during the experiments. The low density in this range qualifies as free flow speed (i.e. walking speed in the absence of crowd). The average speeds of the empirical study and the two simulations are compared in Table . The standard deviations are all small, showing that the speed of pedestrians have had very small deviation from the average value (speed control has been strict). The

-3.1 degree slope is significant in the table because it is the extremum of the hiking function where the maximum speed happens.

Table 1: Simulation of slope for very low density crowds (density ≈ 0.1 ped./m$^2$)

| Slope (Degrees) | Hiking Empirical Function Speed | Simulation (Hiking Function Free Flow Speed) | | | Simulation (Weidmann Free Flow Speed) | | |
|---|---|---|---|---|---|---|---|
| | | Avg. Density | Avg. Speed | St. Dev. Speed | Avg. Density | Avg. Speed | St. Dev. Speed |
| **-40** | *0.1472* | 0.09204 | *0.14691* | 0.00618 | 0.09854 | *0.14493* | 0.00863 |
| **-30** | *0.3054* | 0.07914 | *0.29521* | 0.00829 | 0.08866 | *0.29278* | 0.01150 |
| **-20** | *0.5888* | 0.08570 | *0.59157* | 0.02344 | 0.08522 | *0.54089* | 0.02057 |
| **-10** | *1.0944* | 0.08454 | *1.09427* | 0.03855 | 0.08395 | *1.03934* | 0.03422 |
| **-3.1** | *1.6971* | 0.08249 | *1.69550* | 0.03887 | 0.08268 | *1.59223* | 0.03788 |
| **0** | *1.4211* | 0.08464 | *1.39563* | 0.04053 | 0.08235 | *1.33816* | 0.04290 |
| **10** | *0.7833* | 0.09489 | *0.79268* | 0.02651 | 0.08581 | *0.74348* | 0.01803 |
| **20** | *0.4554* | 0.08856 | *0.44449* | 0.01205 | 0.09335 | *0.43957* | 0.01753 |
| **30** | *0.2111* | 0.10310 | *0.19695* | 0.00311 | 0.10150 | *0.19570* | 0.00574 |
| **40** | *0.1143* | 0.12906 | *0.09854* | 0.00162 | 0.09314 | *0.09872* | 0.00155 |

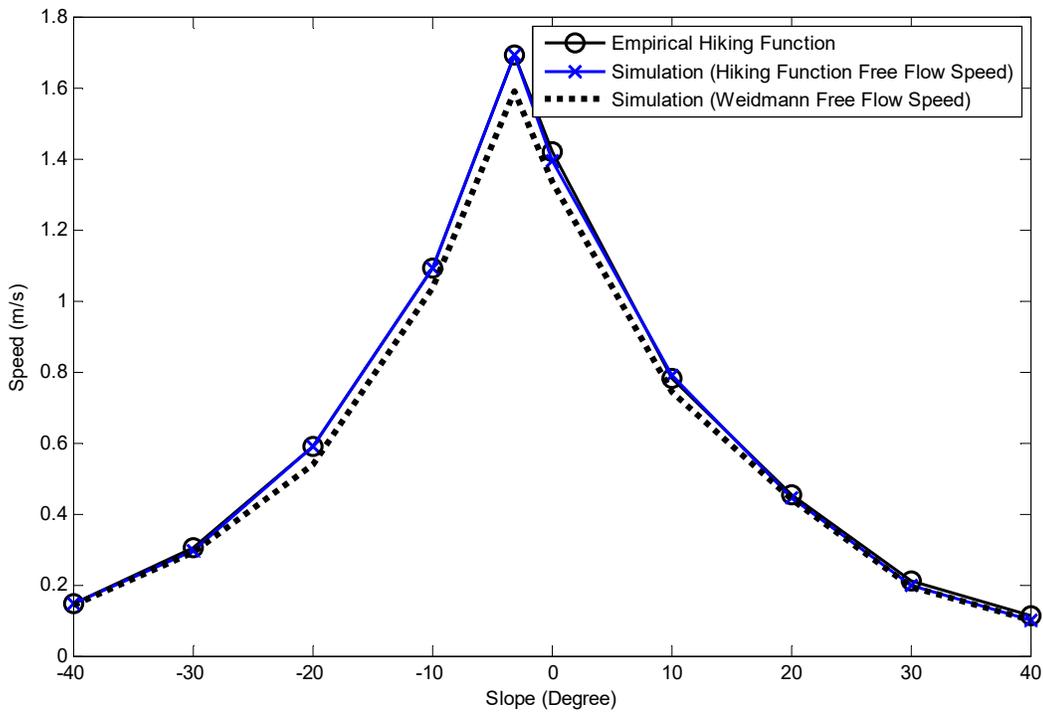

Figure 4: Comparison of slope simulations with the empirical data

Figure shows the comparison in graph form. It can be observed that the simulation has accurately reproduced the empirical data (the two graphs are overlapped) when the free flow speed of the simulation was set equal to the free flow speed of the hiking function (i.e. V$_{(slope=0)}$ = 1.4211 m/s). If the free flow speed is different from the hiking function, as in the case of Weidmann's empirical graph (i.e. V$_{Weidmann(density=0)}$ = 1.344 m/s), the simulated graph will be slightly shifted. In Figure , the dotted line represents the simulation of slope with Weidmann's free flow speed. It can be observed that the graph is similar to the hiking function but slightly shifted downwards.

## 4.1 Effect of Slope on Speed-density Graphs

Now that pedestrian movements on slope have successfully been simulated for low density, more experiments are performed for crowds with higher densities. For the experiment of this section, the crowd movements are simulated on different slopes and the results are compared to the Weidmann's graph (i.e. slope = 0). Notice that in the simulations, Weidmann empirical graph is used for speed-density look up and pedestrian speed control mechanism. At the same time the hiking function is used to adjust the speed values to different slopes. Table shows the speed-density simulation results for different slopes.

Table 2: Simulation speed-density relations for different slopes

| Density (ped/m$^2$) / Slope (degrees) | 0-0.5 | 0.5-1.0 | 1.0-1.5 | 1.5-2.0 | 2.0-2.5 | 2.5-3.0 | 3.0-3.5 | 3.5-4.0 | 4.0-4.5 |
|---|---|---|---|---|---|---|---|---|---|
| -40  | 0.13  | 0.108 | 0.083 | 0.068 | 0.05  | 0.047 | 0.036 | 0.013 | 0.001 |
| -30  | 0.268 | 0.24  | 0.188 | 0.13  | 0.1   | 0.073 | 0.048 | 0.046 | 0.043 |
| -20  | 0.539 | 0.458 | 0.397 | 0.272 | 0.192 | 0.142 | 0.097 | 0.067 | 0.045 |
| -10  | 0.973 | 0.845 | 0.746 | 0.485 | 0.38  | 0.272 | 0.198 | 0.132 | 0.094 |
| -3.1 | 1.517 | 1.319 | 1.121 | 0.785 | 0.578 | 0.441 | 0.303 | 0.222 | 0.169 |
| 0    | 1.277 | 1.122 | 0.877 | 0.653 | 0.493 | 0.333 | 0.248 | 0.186 | 0.176 |
| 10   | 0.694 | 0.597 | 0.505 | 0.368 | 0.273 | 0.194 | 0.139 | 0.095 | 0.080 |
| 20   | 0.397 | 0.352 | 0.280 | 0.220 | 0.149 | 0.110 | 0.071 | 0.050 | 0.042 |
| 30   | 0.194 | 0.154 | 0.122 | 0.099 | 0.076 | 0.050 | 0.047 | 0.047 | 0.028 |
| 40   | 0.097 | 0.088 | 0.058 | 0.049 | 0.045 | 0.028 | 0.001 | 0.001 | 0.001 |

In order to visualize the changes happening to Weidmann's graph on slope, the speed-density graphs for slopes of -20, -3.1, 0 and 20 degrees (Figure ) have been produced. Notice that the 0-degree graph represents the Weidmann's data. For slopes of different values, the Weidmann's graph is not available and the simulations represent the prediction of what would happen to the graph on slopes of different values. It can be observed that the graph for 0-degree slope (Weidmann's graph) has been shifted upwards or downwards depending on the slope angle.

In the experiments of previous section, the simulation results matched the empirical hiking function with high accuracy. The slope model accuracy was therefore confirmed for low density crowds. Since no speed-density graph is available for slope, the model was used to predict what changes may happen to the existing speed-density graphs (e.g. Weidmann) on slopes with different steepness.

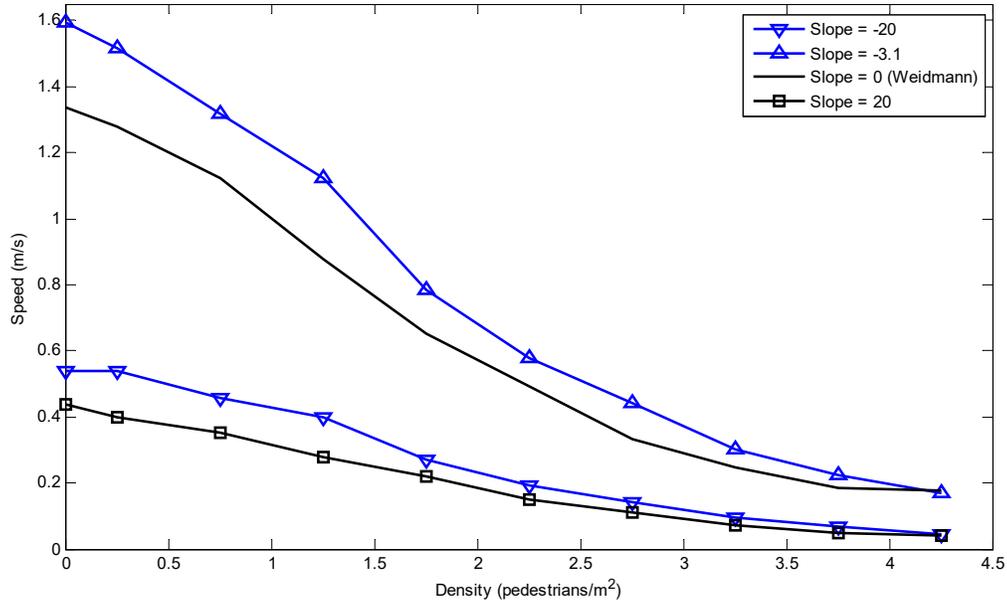

Figure 5: Simulation speed-density graphs for different slopes vs. Weidmann's graph

# 5   Conclusion

Accurate simulation of the crowd movements is important because decisions that might affect the safety and performance of a crowd are taken using their results. In conventional cellular automata models, each pedestrian occupies one cell and cells are of the size of a typical human body. The large size of the cells and the discrete nature of the model dictates a few limited possible speed values for the pedestrians. Fine grid cellular automata on the other hand allows much more speed values. The model can therefore produce more accurate results for sloped floors where the pedestrian speeds depend on the angle of slope. In this article, the fine grid cellular automata model was modified and improved to allow the simulation of crowd movements on sloped surfaces with different degrees of slope and different orientations. The experiments confirmed that the model can achieve accurate results for low density crowds. The model was also used to predict how speed-density graphs may look like on slopes with different steepness. The model was flexible enough to use different empirical density-speed diagrams.